\documentclass[a4paper,12pt]{article}
\pdfoutput=1
\usepackage{graphicx, rotating}
\usepackage{hyperref}
\usepackage{slashed}
\usepackage{ifpdf}
\ifx\pdfoutput\undefined
\usepackage[dvips,bookmarks=false]{hyperref}	% This is for arXiv.org
\else
\usepackage{hyperref}	% This is for pdftex
\fi
\hypersetup{colorlinks,bookmarksopen,bookmarksnumbered,citecolor=verdes,
linkcolor=blus,pdfstartview=FitH,urlcolor=rossos}
\def\hhref#1{\href{http://arxiv.org/abs/#1}{#1}} % in bibliography
\usepackage{amsmath}
\usepackage{slashed}

\newcommand{\beq}{\begin{equation}}
\newcommand{\eeq}{\end{equation}}
\newcommand{\fig}[1]{~\ref{fig:#1}}

\newcount\Mac  \Mac=1  % devo mettere Mac=1 se sto lavorando sul file Mac
\newcommand{\ifMac}[2]{\ifnum\Mac=1 #1 \else #2 \fi}
\def\putps(#1,#2)(#3,#4)#5#6{\ifnum\Mac=1 \put(#1,#2){\special{picture #5}}
\else  \put(#3,#4){\includegraphics{#6}} \fi}

\newcommand{\One}{\hbox{1\kern-.24em I}}

\newcommand{\GeV}{\,{\rm GeV}}

\newcommand{\lascia}[1]{}
\makeatletter
\def\art{\@ifnextchar[{\eart}{\oart}}
\def\eart[#1]#2#3#4#5#6{{\rm #2}, {#3 #4} {\rm (#6) #5} [arXiv:{\hhref{#1}}]}
\def\hepart[#1]#2{{\rm #2, arXiv:\hhref{#1}}}
\newcommand{\oart}[5]{{\rm #1}, {#2 #3} {\rm (#5) #4}}

\newcounter{alphaequation}[equation]
\def\thealphaequation{\theequation\hbox to
0.6em{\hfil\alph{alphaequation}\hfil}}
\def\eqnsystem#1{
\def\@eqnnum{{\rm (\thealphaequation)}}
\def\@@eqncr{\let\@tempa\relax \ifcase\@eqcnt \def\@tempa{& & &} \or
  \def\@tempa{& &}\or \def\@tempa{&}\fi\@tempa
  \if@eqnsw\@eqnnum\refstepcounter{alphaequation}\fi
\global\@eqnswtrue\global\@eqcnt=0\cr}
\refstepcounter{equation} \let\@currentlabel\theequation \def\@tempb{#1}
\ifx\@tempb\empty\else\label{#1}\fi
\refstepcounter{alphaequation}
\let\@currentlabel\thealphaequation
\global\@eqnswtrue\global\@eqcnt=0 \tabskip\@centering\let\\=\@eqncr
$$\halign to \displaywidth\bgroup \@eqnsel\hskip\@centering
$\displaystyle\tabskip\z@{##}$&\global\@eqcnt\@ne
\hskip2\arraycolsep\hfil${##}$\hfil& \global\@eqcnt\tw@\hskip2\arraycolsep
$\displaystyle\tabskip\z@{##}$\hfil
\tabskip\@centering&\llap{##}\tabskip\z@\cr}

\def\endeqnsystem{\@@eqncr\egroup$$\global\@ignoretrue} \makeatother

\def\circa#1{\,\raise.3ex\hbox{$#1$\kern-.75em\lower1ex\hbox{$\sim$}}\,}

\usepackage{multicol}
\usepackage{color}
\definecolor{rosso}{cmyk}{0,1,1,0.4}
\definecolor{rossos}{cmyk}{0,1,1,0.55}
\definecolor{rossoc}{cmyk}{0,1,1,0.2}
\definecolor{blu}{cmyk}{1,1,0,0.3}
\definecolor{blus}{cmyk}{1,1,0,0.6}
\definecolor{bluc}{cmyk}{1,1,0,0.1}
\definecolor{verde}{cmyk}{0.92,0,0.59,0.25}
\definecolor{verdec}{cmyk}{0.92,0,0.59,0.15}
\definecolor{verdes}{cmyk}{0.92,0,0.59,0.4}
\definecolor{grigio}{cmyk}{0,0,0,0.07}
\definecolor{rosa}{cmyk}{0,0.1,0.1,0.02}
\definecolor{rosino}{cmyk}{0,0.05,0.05,0.02}
\definecolor{rosas}{cmyk}{0,0.3,0.25,0.05}
\definecolor{celeste}{cmyk}{0.1,0,0,0.02}
\definecolor{giallino}{cmyk}{0,0,0.4,0.02}
\definecolor{rosso}{cmyk}{0,1,1,0.4}
\definecolor{rossos}{cmyk}{0,1,1,0.55}
\definecolor{rossoc}{cmyk}{0,1,1,0.2}
\definecolor{blu}{cmyk}{1,1,0,0.3}
\definecolor{bluc}{cmyk}{1,1,0,0.1}
\definecolor{blucc}{cmyk}{0.7,0.5,0,0}
\definecolor{viola}{cmyk}{0,1,0,0.6}
\definecolor{viola2}{cmyk}{0,1,0.2,0.6}
\definecolor{verde}{cmyk}{0.92,0,0.59,0.25}
\definecolor{verdec}{cmyk}{0.92,0,0.59,0.15}
\definecolor{verdes}{cmyk}{0.92,0,0.59,0.4}
\definecolor{verdino}{cmyk}{0.12,0,0.09,0.05}
\definecolor{giallo}{cmyk}{0,0,1,0}
\definecolor{gialloverde}{cmyk}{0.44,0,0.74,0}

\font\tenrsfs=rsfs10 at 12pt

\font\sevenrsfs=rsfs7
\font\fiversfs=rsfs5
\newfam\rsfsfam
\textfont\rsfsfam=\tenrsfs
\scriptfont\rsfsfam=\sevenrsfs
\scriptscriptfont\rsfsfam=\fiversfs
\def\mathscr#1{{\fam\rsfsfam\relax#1}}

\def\beq{\begin{equation}}
\def\eeq{\end{equation}}
\def\bea{\begin{eqnarray}}
\def\eea{\end{eqnarray}}

\begin{document}\hfill
 %IFUP-TH/2011-XX\hfill 
 \centerline{CERN-PH-TH/2012-208}

\color{black}
\vspace{1cm}
\begin{center}
{\LARGE\bf\color{black} Is the resonance at 125 GeV  the Higgs boson? }\\
\bigskip\color{black}\vspace{0.6cm}{
{\large\bf Pier Paolo Giardino$^{a}$, Kristjan Kannike$^{{b,c}}$, \\[3mm]
Martti Raidal$^{c,d,e}$
 {\rm and} Alessandro Strumia$^{a,c}$}
} \\[7mm]
{\it (a) Dipartimento di Fisica dell'Universit{\`a} di Pisa and INFN, Italy}\\[1mm]
{\it  (b) Scuola Normale Superiore and INFN, Piazza dei Cavalieri 7, 56126 Pisa, Italy}\\[1mm]
{\it  (c) National Institute of Chemical Physics and Biophysics, Ravala 10, Tallinn, Estonia}\\[1mm]
 {\it (d) CERN, Theory Division, CH-1211 Geneva 23, Switzerland}\\[1mm]
{\it  (e) Institute of Physics, University of Tartu, Estonia}\\[3mm]
\end{center}
\bigskip
\bigskip
\bigskip
\vspace{1cm}

\centerline{\large\bf\color{blus} Abstract}

\begin{quote}\large
The recently discovered resonance at 125 GeV has properties
remarkably close to those of the Standard Model Higgs boson.
We perform model-independent fits of all presently available data.
The non-standard best-fits found in our previous analyses
remain favored with respect to the SM fit,  mainly but not only because the
$\gamma\gamma$ rate remains above the SM prediction.

%The main question we address is whether the Higgs boson with the mass 125 GeV is the standard model one or is there a room for new physics. 
%We find that new 2012 data points towards the standard model Higgs boson couplings and all potential anomalies in 7 TeV data are reduced in 
%combination. While there still is an excess in the  channel,  the other gauge boson channels $ZZ^*$ and $WW^*$ agree, within
%errors, with the SM expectations, proving that the discovered Higgs boson indeed triggers the electroweak symmetry breaking. 
%

\end{quote}
%\end{abstract}

\newpage

%\tableofcontents

\section{Introduction}

New searches for the Higgs boson~\cite{Englert:1964et,Higgs:1964ia,Higgs:1964pj,Guralnik:1964eu,rev}
based on $5$~fb$^{-1}$  data per experiment collected in 2012 by the Large Hadron Collider (LHC) have been recently 
presented by the ATLAS~\cite{ATLAS} and CMS~\cite{CMS} experiments at CERN.
The excess at 125~GeV that was evident already in the 2011 data has been consistently observed in $\gamma\gamma$, $ZZ^*$, $WW^*$ 
channels by  both experiments. In addition,  CMS presented updated Higgs boson searches also in $b \bar b$ and $\tau \bar \tau$ channels.
As a result, when combining the 7~TeV and 8~TeV data, both experiments separately have reached the sensitivity to the SM-like Higgs 
with a significance  of   $5\sigma$.

One must make sure that the discovered new resonance is, indeed, the Higgs boson that 
induces the electroweak symmetry breaking and gives masses to both the SM vector bosons and to fermions.
The SM has definite predictions for the gauge boson and fermion couplings with the Higgs boson.
Those affect both the Higgs boson production mechanism at the LHC as well as its dominant decay modes.
Fortunately for a Higgs boson mass of $125$~GeV  
the LHC experiments do have sensitivity to test these couplings in all interesting 
final states $\gamma\gamma ,$ $ZZ^*,$ $WW^*,$ $b \bar b$ and $\tau \bar \tau$ taking into account different Higgs boson 
productions mechanisms.

The aim of this paper is to perform a fit to the available  Higgs boson data in order to determine its 
preferred couplings to the SM states as well as to an invisible channel.
Our main goal is to study whether the Higgs boson is SM-like or is there any indication for new physics beyond the SM.
The later possibility is motivated by numerous multi-Higgs boson, supersymmetric, composite Higgs boson, dark matter, exotic scalar, etc models.
In the absence of direct signal of new physics, the Higgs boson couplings might indirectly  indicate a portal to new physics.
To achieve this goal we allow the Higgs boson gauge and Yukawa couplings to be free parameters and modify the 
 the Higgs tree level couplings $hWW,$ $hZZ,$ $hf\bar f$ as well as the loop level processes 
 such as the Higgs production in gluon-gluon fusion $gg\to h$ and Higgs decays to $h\to \gamma\gamma, gg,$ accordingly.
  We also allow for an  invisible branching fraction. 
 
The new LHC data on Higgs boson tree level  decays to $WW^*,$ $ZZ^*$  are in a better agreement with the SM expectations than the 2011 year data, proving that the observed state indeed participates in the electroweak symmetry breaking, and  is likely the Higgs boson. We recall that the 2011
data indicated significant deficit in all $WW^*$ channels in both LHC experiments.
Concerning the loop induced observables, the excess in $h\to \gamma\gamma$ observed in 7~TeV data,  in particular the large  excess in  exclusive di-jet tagged events, decreased in the 8~TeV data. 
 However,  the combination of all data still shows an excess in $h\to \gamma\gamma .$  
 
 While our fits are in general model independent, we demonstrate usefulness of our results for constraining new physics beyond the SM 
 using some well known models as examples. These examples show that already in the present stage of accuracy the LHC data constrains
 models severely.

 \medskip
 
In the next section we proceed along the lines of~\cite{Giardino:2012ww} and briefly present updated results;
for technical details of our fitting procedure 
and motivations of the scenarios we consider we refer the reader to our previous paper~\cite{Giardino:2012ww}.

\newpage

\section{Reconstructing the Higgs boson properties}
\label{res}
\begin{figure}
 $$ \includegraphics[height=6.3cm]{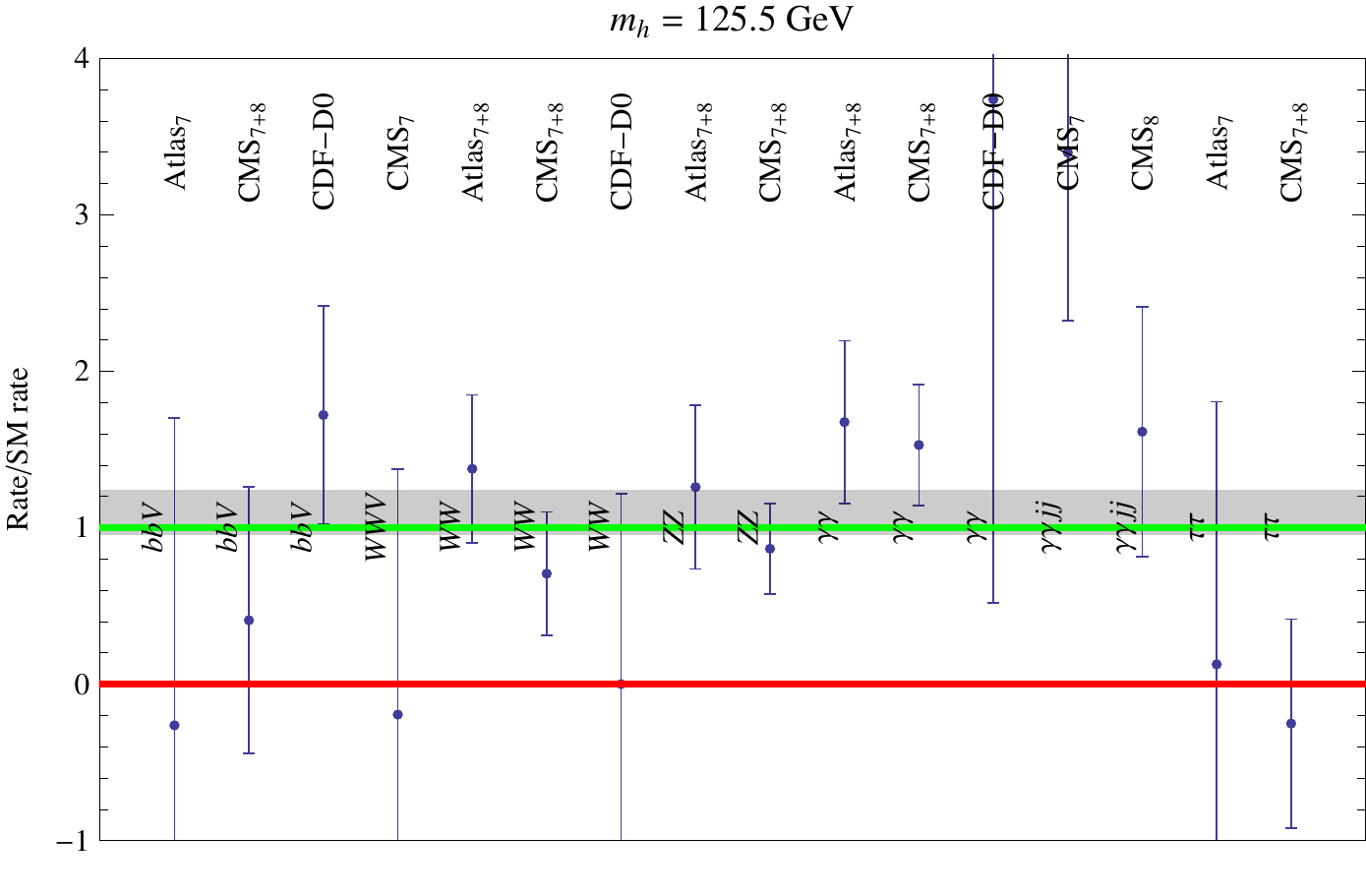}\qquad
 \raisebox{-0.2cm}{\includegraphics[height=6cm]{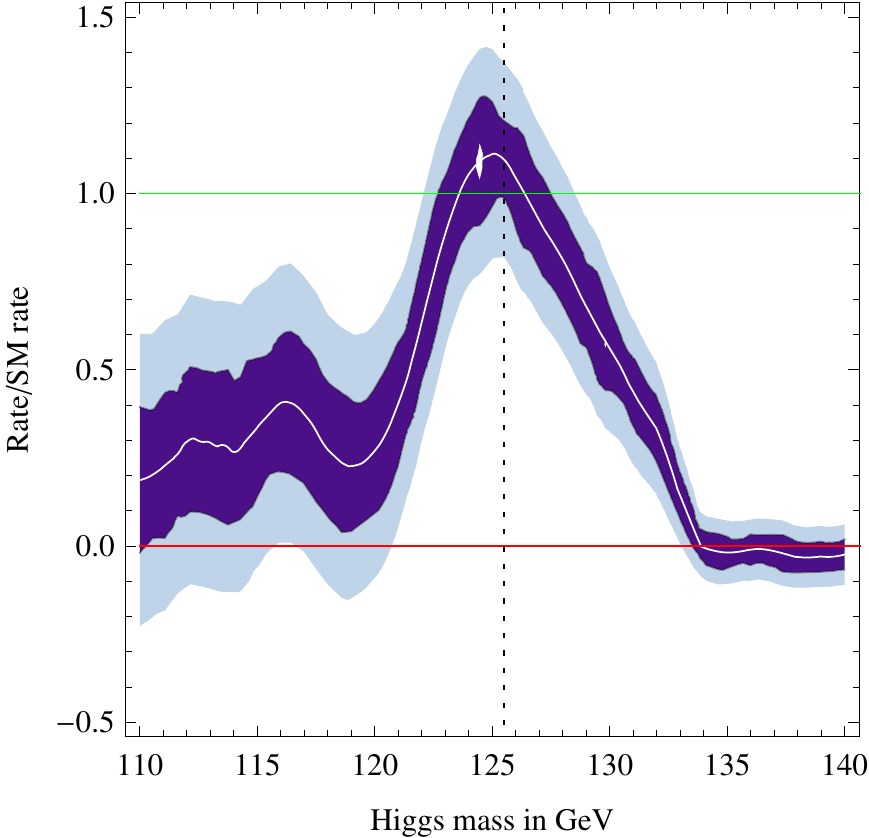}}$$ 
\caption{\em 
{\bf Left}: assuming $m_h=125.5\GeV,$ we show the
measured Higgs boson rates at ATLAS, CMS, CDF, D0 and their average (horizontal gray band at $\pm1\sigma$).
Here 0 (red line) corresponds to no Higgs boson, 1 (green line) to the SM Higgs boson.
{\bf Right}: The Higgs boson rate favored at $1\sigma$ (dark blue) and $2\sigma$ (light blue) in a global SM fit as function of the Higgs boson mass.
\label{fig:data}}
\end{figure}

In the left panel of figure\fig{data}
we summarize all data points~\cite{ATLAS,CMS,comb7,gammagamma,ZZ,WW,bb,tautau,FP,bbTeVatron}
 together with their  $1\sigma$ error-bars.
The grey band shows the $\pm1\sigma$ range for the weighted average of all rates:
\beq \frac{\rm Measured~Higgs~rate}{\rm SM~prediction}=
 1.10\pm 0.15\eeq
It lies along the SM prediction of 1 (green horizontal line) and is 7$\sigma$ away from 0 (red horizontal line). 
Thus the combination of all data favours the existence of Higgs boson with much higher significance than any of the experiments separately.

\subsection{Higgs boson mass}

In the right panel of Fig.\fig{data} we show our approximated combination of all Higgs boson data,
finding that the global best fit for the Higgs boson  mass is
\beq m_h = \left\{\begin{array}{ll}
125.2\pm0.65~\GeV & \hbox{CMS}\\
126.2\pm0.67~\GeV & \hbox{ATLAS}\\
125.5\pm0.54~\GeV & \hbox{combined}
\end{array}\right. \ .
\eeq
The Higgs boson  mass values preferred by the two experiments are compatible, and the uncertainty is so small
that in the subsequent fits we can fix $m_h$ to its combined best-fit value.

\begin{figure}[t]
$$\includegraphics[width=0.44\textwidth]{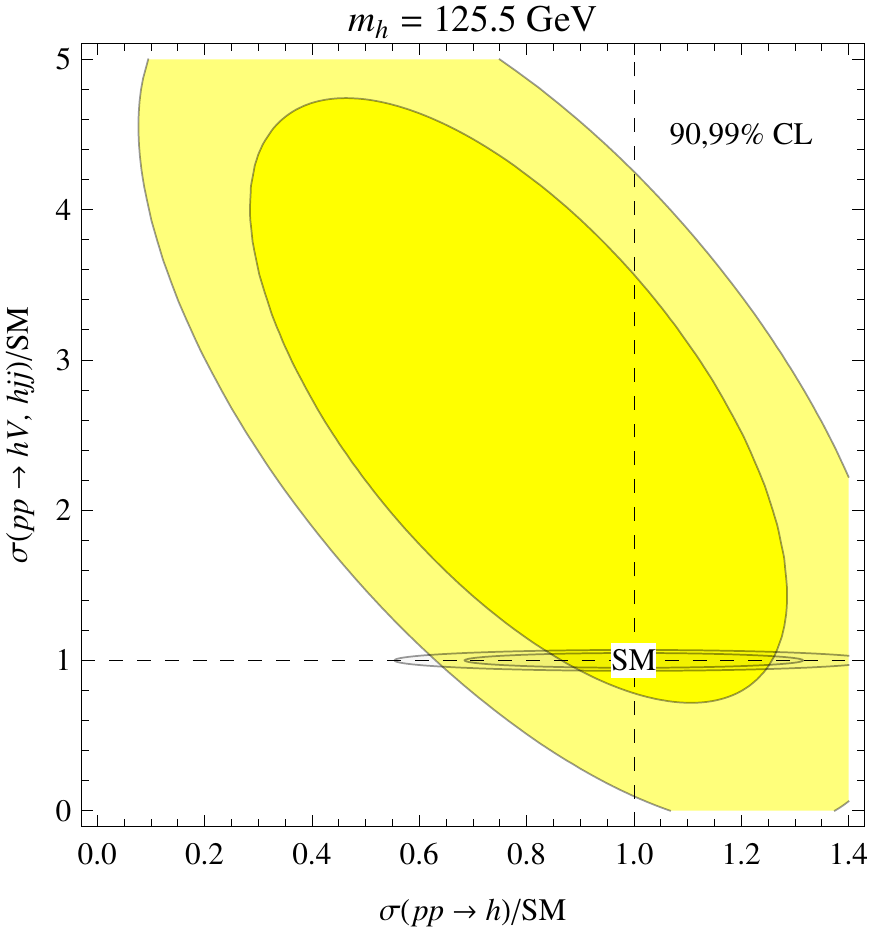}\qquad\includegraphics[width=0.45\textwidth]{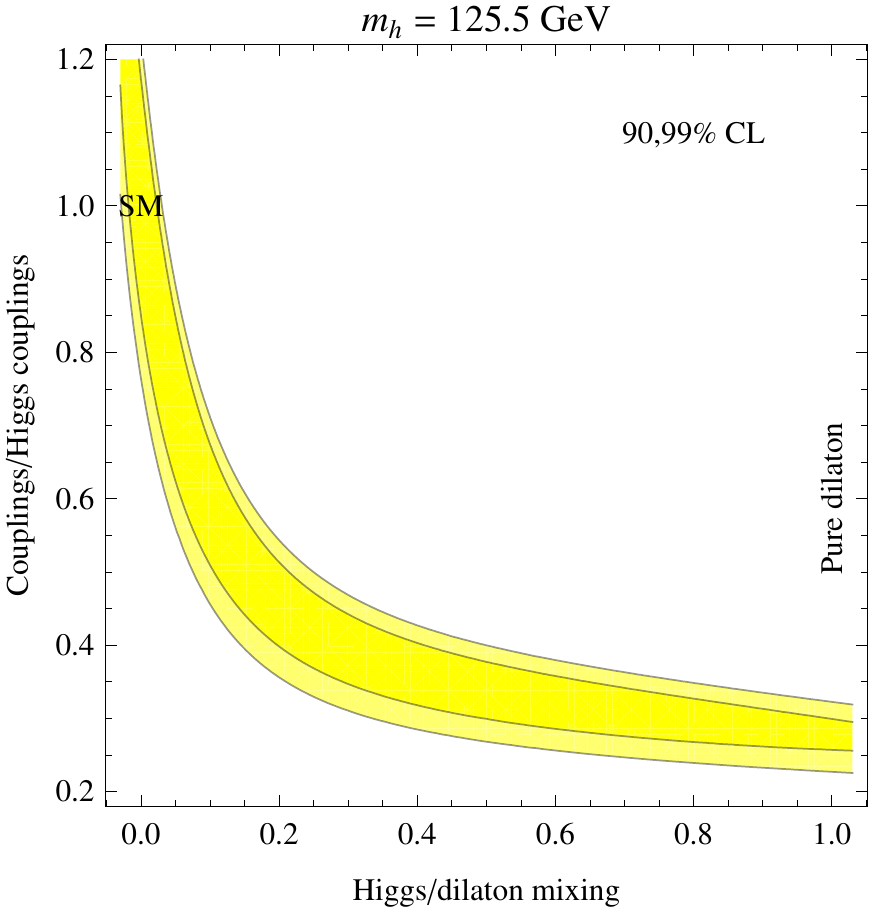}$$ 
\caption{\em {\bf Left}:  determination of SM Higgs boson  production cross-sections compared with
SM theoretical uncertainties (smaller gray ellipses).
{\bf Right}:  fit as function of the Higgs/dilaton mixing (0 corresponds to pure Higgs boson, and 1 to pure dilaton).
\label{fig:sigma}}
\end{figure}

The analysis proceeds along the lines of our previous work~\cite{Giardino:2012ww}
(for similar older fits see~\cite{older}), with the following
modifications: 1) whenever possible we 
use the central values and
the uncertainties on Higgs boson  rates as reported by the experiments,
rather than inferring them from published observed and expected bounds;
2) we take into account uncertainties on the production cross sections:
$\pm14\%$ for $\sigma(gg\to h)$ and for $\sigma(pp\to ht\bar t)$,
$\pm 3\%$ for vector boson fusion, and
$\pm 5\%$ for $\sigma(pp\to Vh)$.
Only the first uncertainty is presently significant, as illustrated in the left panel of Fig.\fig{sigma} where we show
the theoretical and the experimental determination of the production cross sections,
assuming that the resonance at 125.5 GeV is the SM Higgs boson.

\medskip

\begin{figure}[t]
$$\includegraphics[width=0.8\textwidth]{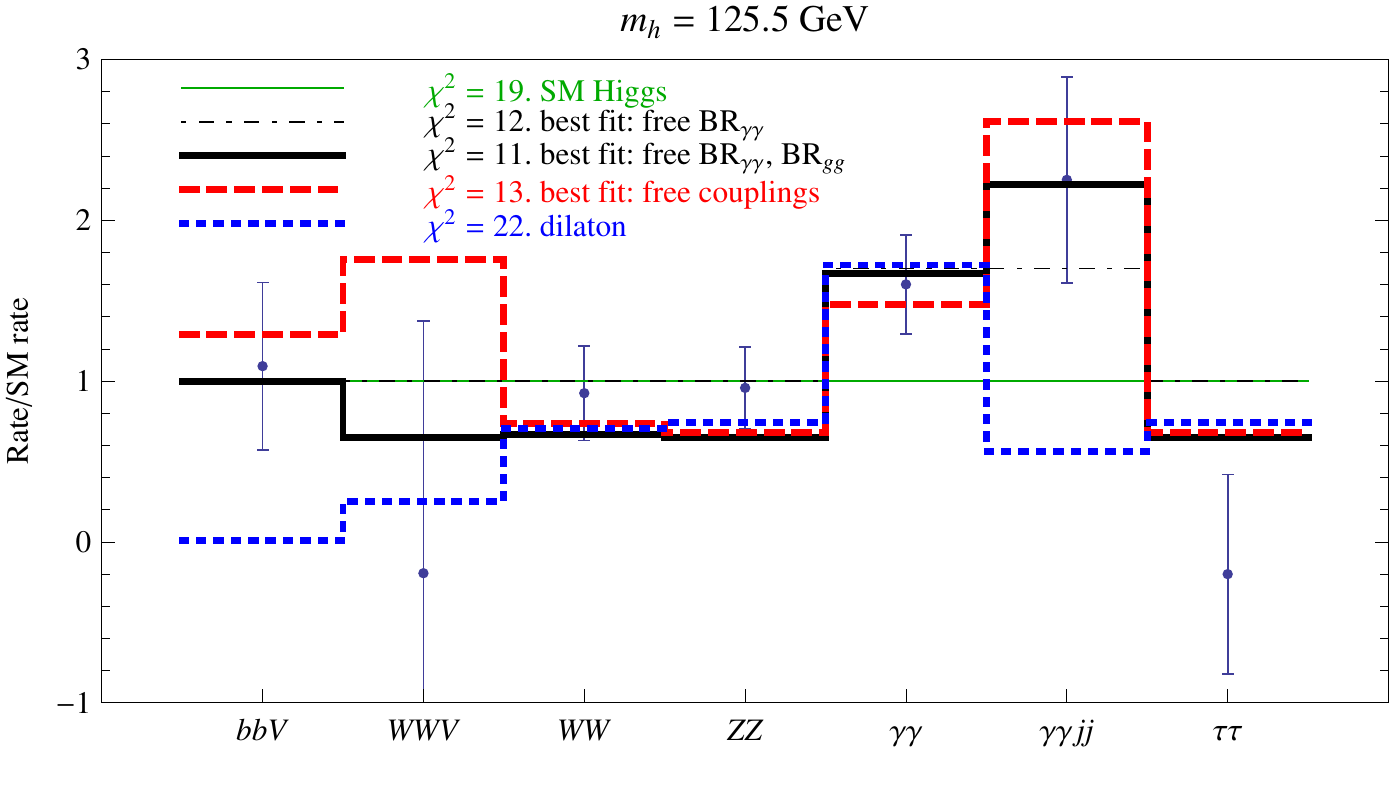}$$
\caption{\em Predictions for the Higgs boson rates in different scenarios: SM, free branching ratios of loop processes,
free couplings, dilaton.
\label{fig:data2}}
\end{figure}

While we always perform a full global fit, in Fig.\fig{data2} we combine the 16 rates into 7 categories according to the final state.
This allows to more clearly see the main features in the data; in particular the $\gamma\gamma$ rate is $1.6\pm0.3$ higher than the SM prediction~\cite{BR,crosssections}, compatibly with the $\gamma\gamma jj$ rate.

In  Fig.\fig{data2}   we also show the rates predicted by a few new physics scenarios.
The SM Higgs boson  (green horizontal line) gives a fit of good quality.
However, a scalar coupled to the trace of the SM energy-momentum tensor
(and thereby called ``dilaton'' or ``radion''~\cite{radion}) gives a fit of overall quality comparable to the SM Higgs boson:
the dilaton fits better the enhanced $\gamma\gamma$ rates but it predicts a
$b\bar b V$ rate below the value preferred by experiments.
Allowing for a mixing between the dilaton and the Higgs boson , all intermediate possibilities give 
 fits with comparable quality, as shown in the right panel of Fig.\fig{sigma}.

Fig.\fig{data2}  also shows that even better fits can be achieved by the
non-standard scenarios discussed that we shall discuss in the following.
%Such scenarios and their motivations have been described in~\cite{Giardino:2012ww}: we here
%mainly update the  results of the fits.

%

\begin{figure}[t]
$$\includegraphics[width=0.45\textwidth]{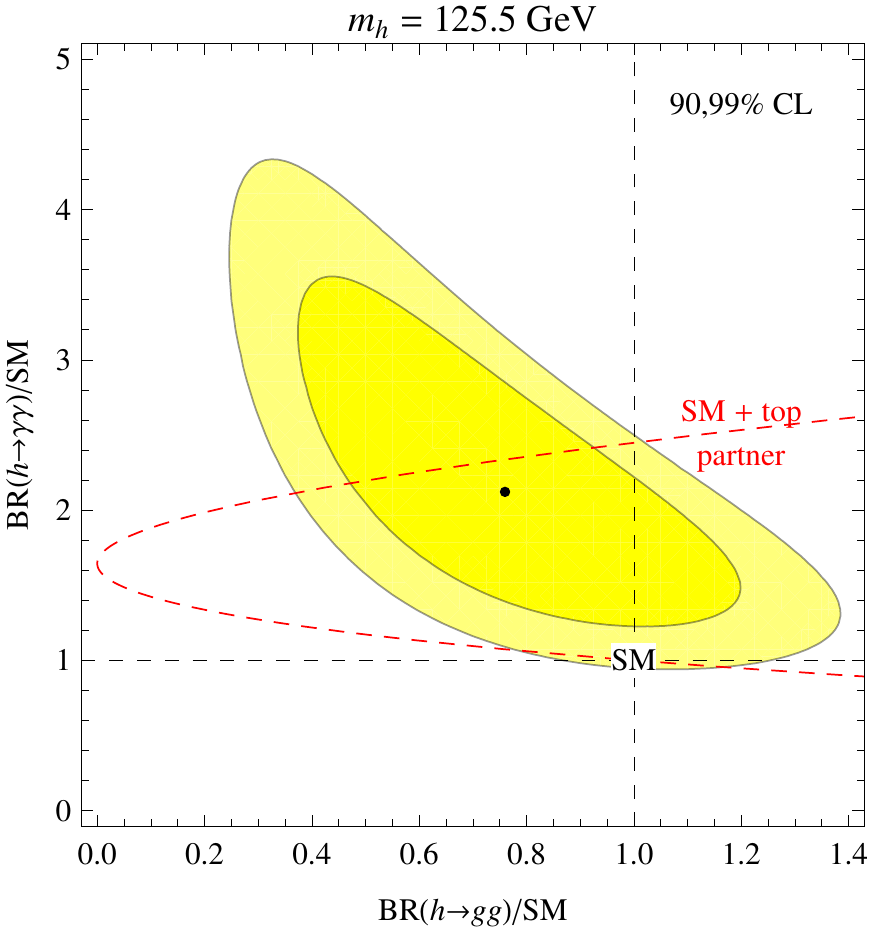}\qquad\includegraphics[width=0.45\textwidth]{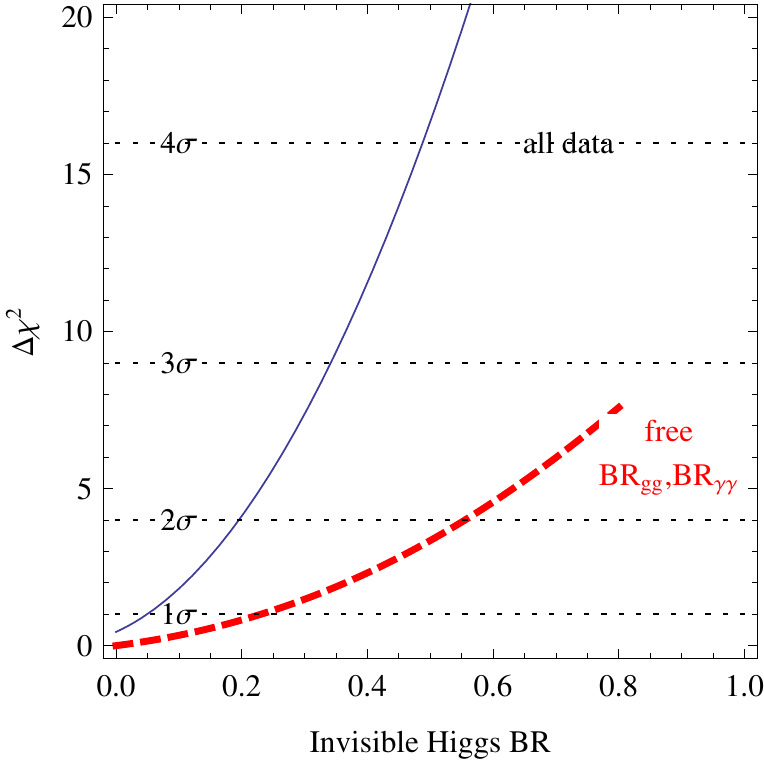}$$ 
\caption{\em {\bf Left}:  fit for the Higgs boson branching fraction to photons and gluons.
The red dashed curve shows the possible effect of extra top partners, such as the stops.
{\bf Right}:  fits for the invisible Higgs boson branching fraction (see section~\ref{inv} for the model assumptions).
\label{fig:fitBR}}
\end{figure}

\subsection{Higgs boson  branching ratios}
In the left panel of Fig.\fig{fitBR} we allow for free values of the rates
that in the SM occur at loop level: $h\to \gamma\gamma$ and
$gg\to h$.  The latter process is related to $h\to gg$ by the well known  Breit-Wigner formula
\beq \sigma(gg\to h)
\stackrel{\Gamma(h)\ll m_h}{\simeq}
\frac{\pi^2}{8m_h} \Gamma(h\to gg) \delta (s-m_h^2),
 \eeq
here written in the relevant narrow-width approximation.
As a consequence we use a common notation BR($h\leftrightarrow  gg$) for those observables when studying their deviations from the SM predictions.
We see that, like in our previous fit, data favour non-standard values
\beq 
\label{BRfit}
\frac{\hbox{BR}(h\leftrightarrow gg)}{\hbox{BR}(h\leftrightarrow gg)_{\rm SM}}\approx 0.6,\qquad
\frac{\hbox{BR}(h\to\gamma\gamma)}{\hbox{BR}(h\to\gamma\gamma)_{\rm SM}}\approx 2.
\eeq
Such a best-fit is shown in Fig.\fig{data2}. It  allows for quite significant reduction of the global $\chi^2$,
in agreement with our previous analysis~\cite{Giardino:2012ww}. When we fix BR($h\leftrightarrow  gg$) to the SM values,
the preferred enhancement of BR($h\to\gamma\gamma$) is is good agreement with the ATLAS and CMS results.

\subsection{Higgs boson  invisible width}\label{inv}
Next, we allow for a Higgs boson  invisible width, as motivated {\it e.g.,}  by models of Dark Matter coupled to the Higgs boson~\cite{Eboli:2000ze}.
We perform two fits.
\begin{enumerate}
\item We just add an additional invisible component to the SM Higgs boson width, 
finding that present data imply 
\beq \hbox{BR}_{\rm inv} = 0\pm0.15,\eeq
as seen  in the right panel of Fig.\fig{fitBR}.
\item In addition to the latter we also  allow for non-standard values of
$h\to \gamma\gamma$ and $h\to gg$, finding a
weaker constraint on BR$_{\rm inv}$, also shown in the right panel of Fig.\fig{fitBR}.
\end{enumerate}
An invisible Higgs boson  width also gives unseen missing-energy signatures, which 
presently provide less stringent constraints~\cite{Fal} on Higgs boson properties than do global fits~\cite{Giardino:2012ww,inv}.

\subsection{Higgs boson  couplings}
Next we extract from data the Higgs boson couplings to vectors and fermions, in order to
test if they agree with the SM predictions.
We recall that the SM predicts a negative interference between the $W$-loop and the top-loop
contributions to $h\to\gamma\gamma$. In general this rate depends on the relative sign of these two contributions
that depends on the relative sign of the gauge and top Yukawa couplings.

\begin{figure}[t]
$$\includegraphics[width=0.45\textwidth]{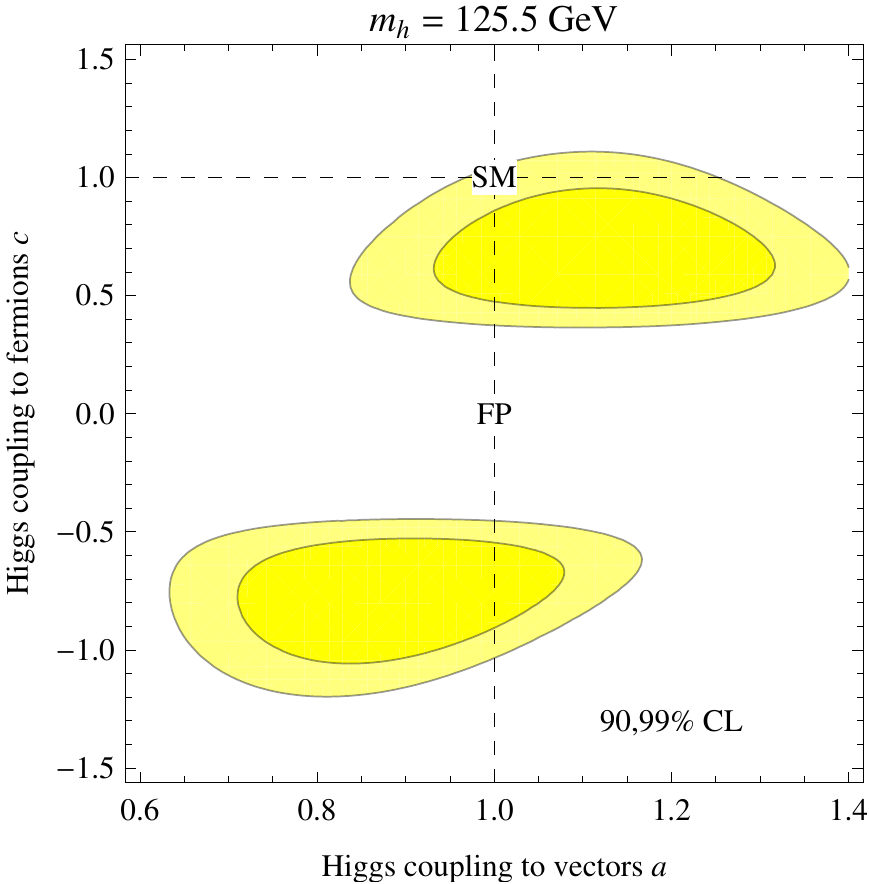}\qquad\includegraphics[width=0.45\textwidth]{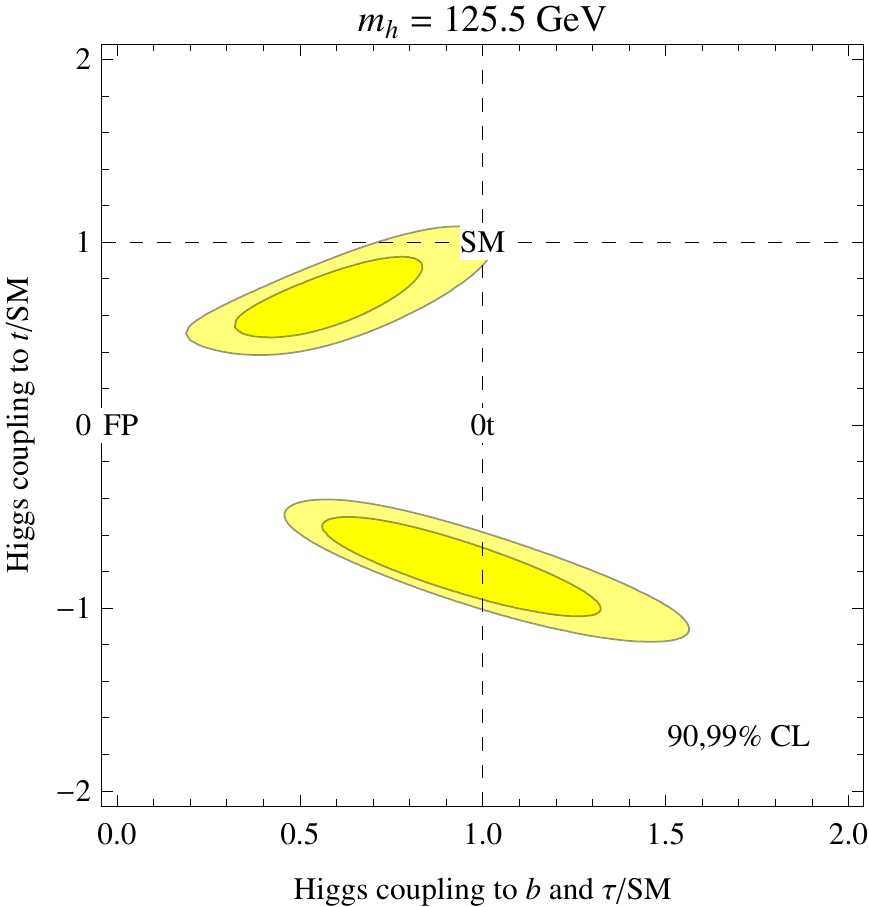}
$$ 
\caption{\em {\bf Left}: fit of the Higgs boson couplings
assuming common rescaling
factors $a$ and $c$ with respect to the SM prediction for couplings to
vector bosons and fermions,  respectively.  {\bf Right}: fit to the $t$-quark and to $b$-quark and $\tau$-lepton Yukawa couplings assuming the SM couplings to gauge bosons.
The point marked as `SM' is the Standard Model;
the point marked as `FP' is the fermiophobic case, and `0t' denotes the top-phobic case.
%Negative values of the top Yukawa coupling are preferred because lead of an enhancement of $h\to \gamma\gamma$.
\label{fig:fitac}}
\end{figure}

In the left panel of Fig.\fig{fitac} we assume a common rescaling of the Higgs boson  coupling to the $W,Z$ bosons and a common rescaling of
the Higgs boson  couplings to all fermions, denoted by $a$ and $c,$ respectively.
We find two preferred solutions, that both allow for an enhancement of $h\to\gamma\gamma$.
The first solution has the Higgs boson  coupling to fermions, thus also to the top and bottom quarks,  reduced with respect to the SM predictions,
thereby reducing the negative interference in $h\to\gamma\gamma$ and increasing its branching fraction. 
This solution prefers somewhat enhanced Higgs boson couplings to vectors that enhances also the $W$-loop contribution to $h\to\gamma\gamma.$
The second solution has  the Higgs boson   coupling to the top quark with opposite sign with respect to
 the SM prediction, thereby making the interference constructive and, again, increasing the branching fraction.
 In this region smaller than SM gauge couplings are preferred.
 
 While the allowed regions in $a$ are quite large and the SM prediction $a=1$ is well within $90\%$ CL region, the Yukawa couplings
 show more non-standard behaviour.
 This is demonstrated in  the right panel of  Fig.\fig{fitac}, where we fix the Higgs boson  coupling to vectors to the SM value
as predicted by gauge invariance and allow the  couplings to fermions to top quark and to bottom quark/tau lepton to vary independently.
We again find the same two best-fit solutions previously discussed. Notice that for positive Yukawa couplings
one significant reduction of all of them is preferred.  Yukawa couplings of order 30\% of the SM values give as good fit as the SM itself.
Although the purely fermiophobic Higgs boson is excluded with high significance, the question of the origin of 
reduced Yukawa couplings and, consequently, the question of 
the new physics contribution to the top/bottom masses remains open.
If the present trends in the LHC data persist, this is one of the most clear signal of physics beyond the SM.

\begin{figure}[t]
$$\includegraphics[width=0.65\textwidth]{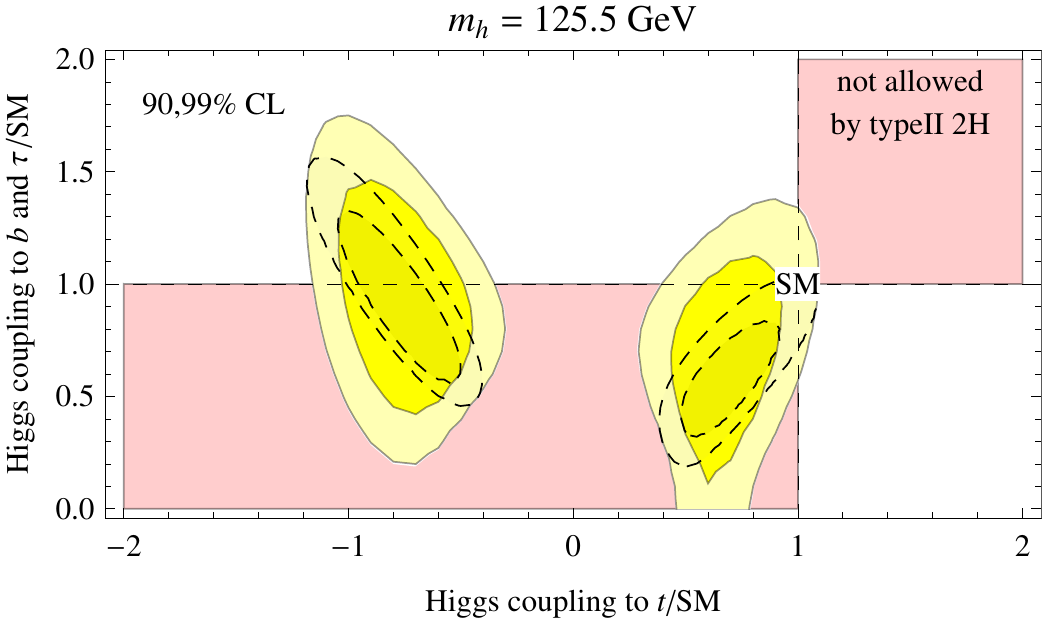}$$ 
\caption{\em Fit in  2 Higgs doublet model either allowing (solid curves) or not (dashed line)
a deviation from the SM in Higgs boson  couplings to vectors.
\label{fig:fitbt}}
\end{figure}

To demonstrate the usefulness of  our fits for constraining models of new physics
 we consider two Higgs doublet model of type II~\cite{Gunion:1989we},  which allows independent 
for a modification of the $t$ coupling, and for a common modification of the 
$b$ and $\tau$ couplings, although  one of them is predicted be reduced and the other enhanced by the model.
We also allow for a modification in the Higgs boson coupling to vectors. The results are presented in Fig.\fig{fitbt}
where we plot our best fits to the Yukawa couplings together with the theoretically forbidden regions of the parameter space.
One sees that in this model the negative Yukawa couplings are strongly preferred. 
For the case of the SM gauge couplings the positive Yukawa region is allowed only at 99\% CL.
This example demonstrates that multi Higgs models may be in difficulties to explain present data and mode exotic
new physics scenarios must be used.

\begin{figure}[t]
$$\includegraphics[width=0.48\textwidth]{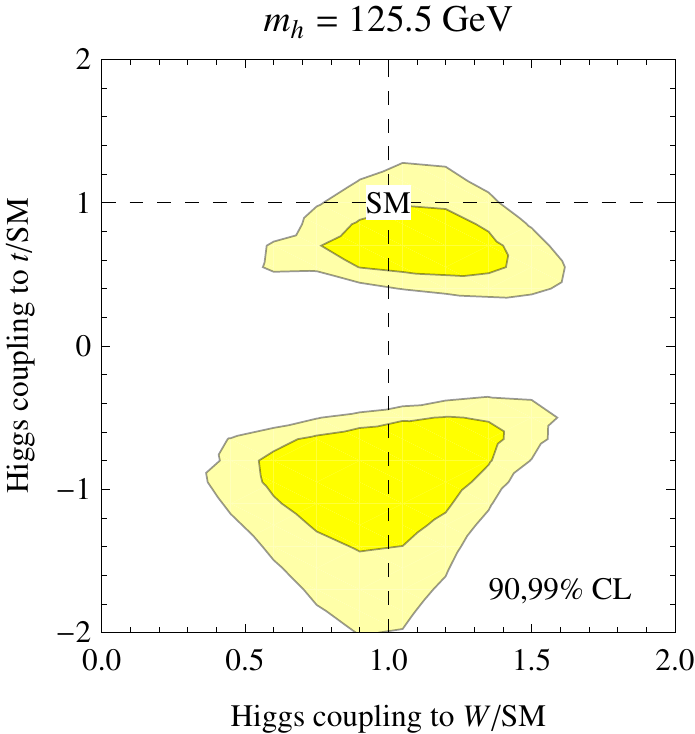}\qquad\includegraphics[width=0.45\textwidth]{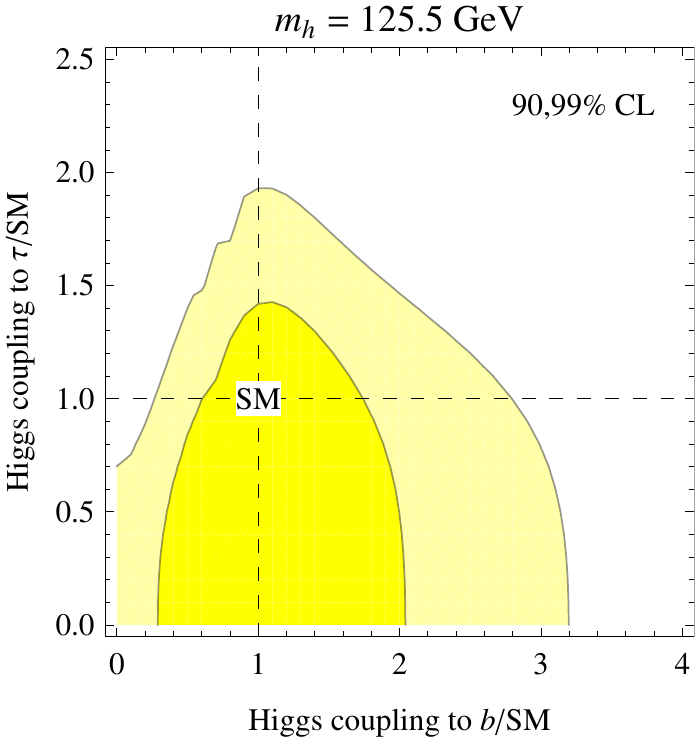}$$ 
\caption{\em Global fit for the Higgs boson couplings to vectors, to the $t$-quark, to the $b$-quark, to the $\tau$ lepton.
All these couplings are freely varied and in each panel we show the $\chi^2$ as function of the  parameters indicated on the axes,
marginalised with respect to all other parameters.  We again assume $m_h=125.5\GeV$ and find
that the best fit presently lies somehow away from the SM prediction,
indicated in the figures as `SM'.
\label{fig:fitR}}
\end{figure}

Finally, in Fig.\fig{fitR}
we allow four Higgs boson  couplings, the ones to gauge bosons, to  top quark, to bottom quark and to tau lepton  to
vary independently. The global fit shows no preference for non-standard gauge couplings,
we once again find the two solutions for Yukawas. Notice that present data are not sensitive to
the Higgs boson  coupling to the $\tau$.
While a tau-phobic Higgs boson  is still allowed (and actually mildly favored),
present data significantly disfavor the pure fermio-phobic or top-phobic or bottom-phobic Higgs boson.
As before, the top Yukawa coupling is the most constrained one and shows significant preference for non-standard values.
More data should confirm that new physics beyond the SM is discovered indirectly due to non-standard Yukawa couplings.

\section{Conclusions}
\label{concl}
The new particle with mass $125.5\pm 0.5$~GeV discovered at the LHC looks like the Higgs boson.
We performed a fit to all available collider data in order to test its couplings.
We find that the couplings to the $W$ and the $Z$ are in reasonable agreement with the SM Higgs boson  expectations,
suggesting that the discovered state is, indeed, the Higgs boson.
However,  the excess in $\gamma\gamma$ indicates
potential non-standard physics in the loop level process  $h\to \gamma\gamma$ (see e.g.~\cite{CW}).
Combining all $\gamma\gamma$ channels and all experiments, this enhancement is at the $2.5\sigma$ level.

As long as this excess persists, it can be fitted by a non-standard (possibly negative) 
Yukawa couplings of the Higgs boson to the top quark, or explained by
 new particles contributions to the loop level process $h\to \gamma\gamma$ 
 and maybe $gg\to h$.
Indeed, allowing for a reduction of $gg\to h$ further improves the global fit.

We considered two main classes of scenarios: a) modified $h\to\gamma\gamma,gg$, which can
be obtained by effective operators of the form $HH^\dagger F_{\mu\nu}^2$;
b) modified Higgs boson  couplings to tops and other fermions, which can be obtained by effective operators of
the form $QUHH^\dagger H$.
In case the anomalies will persist, it will be interesting to explore observables, like $\sigma(gg\to ht \bar{t})$, that
can discriminate among them.

Our general fits were illustrated with some example models. While models with reduced Yukawa couplings 
were used to improve the fit, dilaton (or radion) scenarios and two Higgs doublet model of type II are shown to be 
well constrained already with the present data.

%Assuming no new physics particles in the loops, the best fit indicates partially fermiophobic Higgs boson with negative Yukawa couplings.
%Alternatively, the excess can be explained with new physics contributions to the loop level processes. 
%Since both LHC experiments have collected more than 10/fb data, and since the excess is quite significant,
%it requires lot more statistics to prove that it is an upward statistical fluctuation. New physics in Higgs couplings is still an viable option.

\medskip

\paragraph{Acknowledgement} 
This work was supported by the ESF grants  8090, 8943, MTT8, MTT60, MJD140 by the recurrent financing SF0690030s09 project
and by  the European Union through the European Regional Development Fund.

\footnotesize

\begin{multicols}{2}

\end{multicols}
\end{document}